\documentclass[pra,aps,superscriptaddress]{revtex4}
\usepackage[dvips]{graphicx}
\usepackage{ae}
\usepackage{physics}
\usepackage[T1]{fontenc}
\usepackage[ansinew]{inputenc}
\usepackage{amsmath}
\usepackage{amssymb}
\usepackage{graphicx}
\usepackage{color}
\usepackage[colorlinks]{hyperref}
\usepackage{lscape}
\usepackage{mathtools}
\usepackage{csquotes}
\usepackage{subfig}
\usepackage{xstring}
\NewDocumentCommand{\zoom}{ m m }{%
	\IfEqCase{#1}{%
		{c}{\resizebox{\columnwidth}{!}{$#2$}}
		{t}{\resizebox{\textwidth}{!}{$#2$}}
		{p}{\resizebox{\paperwidth}{!}{$#2$}}
		{l}{\resizebox{\linewidth}{!}{$#2$}}
	}[\scalebox{#1}{$#2$}]
}

\begin{document}
\title{Entanglement concentration via measurement:- role of imaginarity}

\author{Indranil Biswas}
\email{indranilbiswas74@gmail.com}
\affiliation{Department of Applied Mathematics, University of Calcutta, 92, A.P.C. Road, Kolkata- 700009, India}
\author{Subrata Bera}
\email{98subratabera@gmail.com}
\affiliation{Department of Applied Mathematics, University of Calcutta, 92, A.P.C. Road, Kolkata- 700009, India}
\author{Ujjwal Sen}
\email{ujjwal@hri.res.in}
\affiliation{Harish-Chandra Research Institute, A CI of Homi Bhabha National Institute, Chhatnag Road, Jhunsi, Prayagraj 211019, India}
\author{Indrani Chattopadhyay}
\email{icappmath@caluniv.ac.in}
\affiliation{Department of Applied Mathematics, University of Calcutta, 92, A.P.C. Road, Kolkata- 700009, India}
\author{Debasis Sarkar}
\email{dsarkar1x@gmail.com, dsappmath@caluniv.ac.in}
\affiliation{Department of Applied Mathematics, University of Calcutta, 92, A.P.C. Road, Kolkata- 700009, India}
\begin{abstract}
The role of complex numbers in quantum theory extends beyond mathematical convenience, having recently been formalized as a resource under the framework of the resource theory of imaginarity. Operationally, imaginarity translates into using fewer resources in optical setups. In this work, we investigate the operational advantage offered by complex-valued measurements in the entanglement of assistance protocol for three-qubit systems. We demonstrate that employing such measurement bases leads to a significant improvement in the concentration of bipartite entanglement with the aid of the third party. We further analyze a modified entanglement swapping protocol and show that a three-qubit complex measurement bases with certain symmetries outperform the standard GHZ-basis. This is also one example where a three-qubit non-maximally entangled basis surpasses a maximally entangled one in generating entanglement. Construction of the basis also addresses the open problems raised in [Phys. Rev. A. \textbf{108}, 022220 (2023)]. As an intriguing application, we show that using this approach in quantum network percolation on a honeycomb lattice reduces the required bond occupation probability by $22.7\%$ and, requirement of entanglement by $10.6\%$ in each bond.
\end{abstract}
\date{\today}
\maketitle
\section{Introduction}
\label{sec1}

Quantum resource theories \cite{1,2} have emerged as a powerful framework in recent years, drawing significant interest for their ability to formally describe and quantify various physical phenomena that are considered useful for quantum technologies. From the early stages of quantum information theory, entanglement \cite{3} has been viewed as one of the most valuable resources, enabling tasks such as teleportation \cite{4}, superdense coding \cite{5}, and quantum cryptography \cite{6}--tasks that are otherwise unattainable using classical means. In the context of any resource theory, it is essential to identify two key components: the set of states that are considered ineffective or useless for the task at hand, known as free states, and a set of operations known as free operations, which are allowed within the theory. These operations should be capable of manipulating resourceful states but must not be able to generate resource from free states.

In the case of entanglement, separable states are typically considered free, and the operations most commonly regarded as free are Local Operations and Classical Communications (LOCC) \cite{7}. LOCC is regarded as the most physically implementable set of operations and forms the foundation of many quantum protocols. However, alternative classes of free operations such as separable operations (SEP) \cite{8,9}, positive partial transpose (PPT) operations \cite{10}, non-entangling operations, etc., have gained attention for their utility in exploring the boundaries of quantum theory and for deepening our understanding of the structure of entanglement. Beyond entanglement, resource theories have been constructed for other quantum features like coherence \cite{11,12}, asymmetry, thermodynamics and imaginarity, enriching the theoretical toolbox for advancing quantum technologies.

It is worth noting that most of the effort in developing quantum resource theories has focused on the classification and manipulation of quantum states. This approach assumes that resource states are the carriers of non-classical power in protocols. For example, in the standard teleportation protocol, the resource is a maximally entangled state shared between two distant parties, and the operations are LOCC. However, a crucial yet often overlooked component of such protocols is the role played by resource in measurements \cite{13}. In teleportation, the protocol cannot be completed unless the input party performs a standard Bell State Measurement (BSM) locally. This underscores that entanglement in measurements--not just in states, plays an essential role in unlocking quantum advantages.

Traditionally, BSM has been employed as a standard measurement tool in various quantum communication protocols, including teleportation, superdense coding, and entanglement swapping \cite{14,15}. Yet recent research has revealed that this standard may not always be optimal. Specifically, there is growing evidence that certain non-maximally entangled measurements can outperform BSM in generating stronger non-classical correlations. Gisin has constructed a new elegant joint measurement (EJM) \cite{13,16,17,18} basis by considering a different symmetry than the BSM. This has proven to be a more powerful resource compared to BSM for detecting violations of bilocality \cite{19,20,21}. Santo \emph{et} \emph{al.} \cite{22} presented a more rigorous and general study on two-qubit joint measurements, encompassing BSM and EJM altogether. Wei \emph{et} \emph{al.} \cite{23} have used the mathematical concept of equiangular tight frames to introduce a set of two-qubit measurements that are non-projective in nature. This has spurred interest in systematically characterizing measurements as resources, paralleling the established theories for states.

One interesting observation we make in all these constructions is that unlike BSM, all the measurement basis comprises of complex coefficients. The necessity of complex numbers in quantum mechanics has long been accepted, but their operational advantage has only recently been recognized as a quantifiable resource \cite{24}. This has led to the development of the resource theory of imaginarity. Although the measurement bases mentioned above are all non-maximally entangled, from the perspective of resource theory of imaginarity, these bases are more resourceful than the standard BSM. In this work, we primarily focus on this less-explored but increasingly relevant resource: imaginarity--the use of imaginary coefficients--in measurement bases. Building on this, we explore how complex numbers in measurement bases can lead to practical advantages in quantum information tasks.

Our analysis begins with entanglement of assistance \cite{25} protocols involving three-qubit pure states, where one party assists two others in concentrating entanglement via local measurements.
By analyzing three-qubit slice states, we demonstrate that imaginary components in the measurement basis can significantly enhance the ability to localize entanglement between two parties and also briefly describe implementing in linear optical setups. Furthermore, we extend our study to a generalized three-qubit entanglement swapping protocol. In the standard version of this protocol, three pairs of maximally entangled states are shared between three pairs of parties $A\leftrightarrow B,\;A\leftrightarrow C$ and $A\leftrightarrow D$, and the goal is to create genuine three-qubit entanglement by performing a three-qubit joint measurement. The genuinely entangled state can then be converted into a bipartite entangled state among arbitrary pair of parties at a later stage. It is commonly assumed that using the maximally entangled GHZ-basis, a natural generalization of Bell basis comprising of maximally entangled three-qubit states---yields the best performance. However, our findings challenge this assumption. We show that when the initial bipartite entangled states are non-maximally entangled, the GHZ-basis no longer constitutes the optimal choice. Instead, we introduce an entangled basis with imaginary coefficients that can outperform the GHZ-basis, providing higher post-measurement entanglement between the target pair.

Finally, we illustrate a compelling application of this phenomenon in the context of entanglement percolation \cite{26}--a method used to establish long-range entanglement across a network. By incorporating the complex-valued measurement basis, we show that the bond occupation probability required in a honeycomb lattice can be reduced by as much as $22.7\%$, and requirement of ebits in each bond by $10.6\%$. This striking result highlights the operational significance of imaginarity and opens up new directions for optimizing quantum networks and distributed quantum computation.\\
\section{Entanglement of Assistance (EoA) and Optimal Decoupling}
\label{sec2}
One of the essential uses of a genuinely entangled three-qubit state $\ket{\psi}_{ABC}$, shared among three parties $A,\; B$ and $C$ is to create an entangled bipatite state at a later time. The amount of entanglement that can be locally concentrated between two parties, say $A$ and $B$ with the local assistance of $C$ is given by the entanglement of assistance (EoA) \cite{25} of $\ket{\psi}_{ABC}$ with respect to $C$.
Formally the EoA of $\ket{\psi}_{ABC}$ with respect to $C$ is denoted by $E^{A|B}$ and is defined by:
\begin{equation}
\label{eq1}
E^{A|B} (\ket{\psi}_{ABC}) \vcentcolon = \max \sum_i p_i E(\ket{\phi_i}_{AB})
\end{equation}
where the maximization is over all decomposition of $\Trace_C ({\ket{\psi}_{ABC}}{\bra{\psi}})=\sum_i p_i \ket{\phi_i}_{AB}\bra{\phi_i}$ and the the quantity $E(\cdot)$ in the right hand side denotes any relevant bipartite entanglement measure. The EoA also happens to be upper bounded by a very natural condition:
\begin{equation}
\label{eq2}
    E^{A|B} (\ket{\psi}_{ABC}) \leqslant \min \{ E^{A|BC}(\ket{\psi}_{ABC}), E^{B|AC}(\ket{\psi}_{ABC}) \}
\end{equation}
where $E^{A|BC}(\cdot)$ and $E^{B|AC}(\cdot)$ denotes the entanglement of the bipartite states when parties $B\leftrightarrow C$ and $A\leftrightarrow C$ are respectively put together.

Whether the inequality (\ref{eq2}) is eventually becomes tight for some bipartite entanglement measure is an intriguing question. Pollock \emph{et} \emph{al.} \cite{27} found that if the bipartite entanglement is measured by the probability of the state being converted into a singlet \cite{28}, then the inequality becomes an equality and we denote this measure $E_2(\cdot)$ such that:\\
\begin{equation}
\label{eq3}
    E_2^{A|B} (\ket{\psi}_{ABC}) = \min \{ E_2^{A|BC}(\ket{\psi}_{ABC}), E_2^{B|AC}(\ket{\psi}_{ABC}) \}
\end{equation}
for every pure three-qubit state $\ket{\psi}_{ABC}$. The singlet conversion probability ($SCP$) of $\ket{\psi}_{ABC}$ in any of the bipartitions say, $A|BC$ is given by two times the smallest eigenvalue of the state $\Trace_{BC}(\ket{\psi}_{A|BC})$. The relation (\ref{eq3}) essentially means that $C$ could locally concentrate the entanglement between $A$ and $B$, and can decouple himself from the system without losing any entanglement in the process.\\
The saturation of the inequality (\ref{eq2}) also implies that $C$ can now find a measurement setup that can optimally concentrate the entanglement between $A$ and $B$. We will make extensive use of this fact throughout the paper to derive our results. In the next section, we present a brief overview of the resource theory of imaginarity, which sets the stage for our main analysis.

\section{Resource Theory of Imaginarity}
\label{sec3}
The motivation for any resource theory \cite{1,2} generally comes from the fact that certain \enquote*{states} are valuable for specific physical tasks. One of the principal reason behind developement of resource theory imaginarity (RTI) \cite{24,29} is the relative ease of creating real states in experimental setups. For instance in polarization encoded photonic systems, we can perform any rotation around y-axis by a single half wave plate. However, to perform a rotation around z-axis, we have to use two additional quarter wave plates. Analogous to the resource theory of coherence, RTI is also basis dependent. Here the free states are the \enquote*{real} states $\ket{\psi}$ such that
\begin{center}
$$
\ket{\psi_r}=\sum_i r_i \ket{i}
$$
\end{center}
where $r_i\in\mathbb{R}\;\forall i$ ($\mathbb{R}$ being the set of all real numbers) satisfying $\sum_i r_i^2=1$, with respect to the computational basis \{$\ket{i}$\}. For mixed states, with respect to the computational basis, a state $\rho$ is real if it can be represented as below:
\begin{center}
$$
\rho = \sum_{j,k}\rho_{jk}\ketbra{j}{k}
$$
\end{center}
such that $\rho_{jk}\in\mathbb{R}$, $\forall j,k$. The resourceful states in RTI are the states having complex coefficients in the chosen basis \{$\ketbra{i}{i}$\}. In particular, a state $\ket{\phi}$ is said to be resourceful with respect to the basis \{$\ketbra{i}{i}$\} if it can be written as:
\begin{center}
$$
\ket{\psi_c}=\sum_i c_i \ket{i}
$$
\end{center}
where $c_i\in\mathbb{C}$ ($\mathbb{C}$ being the set of complex numbers) satisfying $\sum_i \abs{c_i}^2=1$, with at least one $j$ such that $\Im{c_j}\neq 0$.
As in the resource theories of entanglement and coherence, the RTI also has a maximally resourceful state. It is called the maximally imaginary state and is given by
$$
\ket{\hat{+}}=\frac{\ket{0}+\mathrm{i}\ket{1}}{\sqrt{2}}
$$
where $\mathrm{i}=\sqrt{-1}$. Note that the state $\ket{\hat{+}}$ can also be converted into $\ket{\hat{-}}=(\ket{0}-\mathrm{i}\ket{1})/{\sqrt{2}}$ by a real orthogonal matrix. The state $\ket{\hat{-}}$ is infact also maximally resourceful. This is due to the fact that the any \enquote*{real} operation is infact a \enquote*{free} operation under RTI. Let us formally define below what exactly do we mean by \enquote*{real} operation.\\
\textit{Definition 1.-} Any physical quantum operation is represented by a set $\{K_i\}$ of Kraus operators satisfying $\sum_i K_i^\dagger K_i=\mathbb{I}$, $\mathbb{I}$ being the identity operator. A quantum operation is called real iff $\braket{m| K_i |n}\in\mathbb{R}$, $\forall i,m,n$ in the computational basis.\\
Any real operation applied on some imaginary state should not increase the imaginarity of the state. To properly capture this phenomena several imaginarity measures has been recently introduced, viz., Robustness of Imaginarity (RoI) \cite{29}, Geometric Imaginarity \cite{30}, etc. For the present work we will be using Robustness of Imaginarity (RoI). For any state $\rho$, its RoI is defined as:\\
\begin{align*}
\mathcal{R}(\rho)=&\min_\tau\{s\geqslant 0 : \frac{\rho+s\tau}{1+s} \;\text{is a real state}\}\\
=&\frac{1}{2} \|\rho-\rho^\mathsf{T}\|_1
\end{align*}
where $\|M\|_1=\Trace\sqrt{M^\dagger M}$ and $\rho^\mathsf{T}$ denotes the transpose over the state $\rho$. We have used RoI to measure the imaginarity of a qubit imaginary measurement basis.\\
\textit{Definition 2.-} A measurement basis $\mathcal{M}=\{M_i:M_i\geqslant0,\sum_i M_i =\mathbb{I}\}$ is said to be imaginary if there exist at least one $M_k$ such that $\mathcal{R}(M_k/\Trace{M_k})\neq 0$.\\
In the next section we show that imaginary measurement bases play a crucial role in entanglement concentration.
\section{Single-Qubit Measurement}
\label{sec4}
An arbitrary pure three-qubit state $\ket{\psi}_{ABC}$ can be canonically represented by \cite{31}
\begin{equation}
\label{eq4}
\ket{\psi}_{ABC} = \lambda_0 \ket{000} + \lambda_1 \mathrm{e}^{\mathrm{i}\phi} \ket{100} + \lambda_2 \ket{101} + \lambda_3 \ket{110} + \lambda_4\ket{111}
\end{equation}
where $\mathrm{i}=\sqrt{-1}$ and $\phi\in[0, \pi]$. The coefficients $\lambda_i$'s are all non-negative real numbers and satisfy the relation $\sum_{i=0}^4 \lambda_{i}^2 = 1$.
The entanglement present in every reduced system $\rho_{AB},~ \rho_{AC}$ and $\rho_{BC}$ are given by \cite{32}
\begin{equation*}
\mathbf{C}_{AB} = 2\lambda_0\lambda_3,~ \mathbf{C}_{AC} = 2\lambda_0\lambda_2,~  \mathbf{C}_{BC} = 2\abs{\lambda_2\lambda_3 - \mathrm{e}^{\mathrm{i}\phi}\lambda_1\lambda_4}
\end{equation*}
where $\mathbf{C}_{ij}$ denotes the two-qubit concurrence of the state $\rho_{ij}$. Further, the three-tangle of the state (\ref{eq4}) is given by
\begin{equation*}
\tau_{ABC} = 4\lambda_0\lambda_4.
\end{equation*}
	The three-qubit pure states are divided into two SLOCC inequivalent classes \cite{33} : \emph{GHZ class} and \emph{W class}. For every \emph{GHZ} class state, $\tau_{ABC}\neq 0$. But for \emph{W} class states, $\tau_{ABC} = 0$ \cite{34}.\\
In Ref. \cite{27}, the procedure to concentrate entanglement between two parties with the assistance of the third party has been shown. However we ask whether it is possible to find an universal measurement basis or not. For instance, it is well known that $\{\ket{\pm}\}$ where $\ket{\pm}=(1/\sqrt{2})\ket{0\pm 1}\}$, the eigenstates of Pauli $\sigma_x$ is an optimal measurement basis for generalized \emph{GHZ} ($gGHZ$) state : $\lambda_0\ket{000} + \lambda_4\ket{111}$ and does not depend on the particular parameter values. Thus in order to optimally decouple the state from any party, one only needs to know that the state is gGHZ. Apart from this, the generalized \emph{W} ($gW$) state : $\sqrt{x_1}\ket{100} + \sqrt{x_2}\ket{010} + \sqrt{x_3}\ket{001}$ with $x_1+x_2+x_3 = 1$, can also be optimally decoupled using the eigenbasis $\{\ket{0},\ket{1}\}$ of Pauli $\sigma_z$.\\
From the perspective of resource theory of coherence, the basis $\{\ket{\pm}\}$ is a maximally coherent state. As a matter of fact, another maximally coherent basis $\{\ket{\hat{\pm}}\}$ is also an optimal basis for $gGHZ$ state. However, if one considers resource theory of imaginarity, $\ket{\hat{\pm}}$ shows no apparent advantage in this case.\\\\
\textit{Parametric independence for Slice states.-}  Pure three-qubit slice states \cite{35} are \emph{GHZ} class states satisfying the condition $\mathbf{C}_{ij}=0$, $\mathbf{C}_{ik}=0$ and $\mathbf{C}_{jk}\neq 0$, with $i,j,k$ being all distinct. Specfically maximally slice states possess nonlocal properties of significant interest \cite{36}. A slice state with $\mathbf{C}_{AB} = 0,~ \mathbf{C}_{AC} = 0,~  \mathbf{C}_{BC} \neq 0$ is given by
\begin{equation}
\label{eq5}
\ket{\psi_s}=\lambda_0\ket{000} + \lambda_1\ket{100} + \lambda_4\ket{111},
\end{equation}
where $\lambda_0,\lambda_1,\lambda_4$ are non-negative real numbers satisfying $\lambda_0^2+\lambda_1^2+\lambda_4^2=1$.\\
The EoA with respect to  three different parties are given by
\begin{gather*}
E_2^{A|B}(\ket{\psi_s}) = E_2^{A|C}(\ket{\psi_s}) = 1-\sqrt{1-4\lambda_0^2\lambda_4^2}\\
E_2^{B|C}(\ket{\psi_s}) = 2\min\{\lambda_4^2,1-\lambda_4^2\}
\end{gather*}
We find that $E_2^{A|B}(\ket{\psi_s})$ and $E_2^{A|C}(\ket{\psi_s})$ can be locally achieved in a parameter independent way using the real measurement basis $\{\ket{\pm}\}$. However, when it comes to concentrating entanglement between $B$ and $C$, this basis is no longer optimal. Nevertheless to find the measurement basis with real coefficients, we consider the following generic real qubit projective measurement basis:
\begin{align*}
\ket{+k_\alpha} &= \cos\alpha\ket{0} + \sin\alpha\ket{1}\\
\ket{-k_\alpha} &= \sin\alpha\ket{0} - \cos\alpha\ket{1}
\end{align*}
where $\alpha\in[0,\pi/2)$. We find that the optimal real measurement is achieved by choosing $$\alpha=\tan^{-1}\Bigl(\frac{\sqrt{1-\lambda_4^2}+\lambda_1}{\lambda_0}\Bigr).$$
Clearly the optimal measurement basis depend on the specific parameter values of the given slice state. Thus one first needs to perform local state tomography in order to estimate the parameter values and then setup the measurement apparatus. Intriguingly the maximally imaginary basis $\{\ket{\hat{\pm}}\}$ proves to be an optimal basis for this case. For the other two classes of slice states with $\mathbf{C}_{AB} \neq 0$ and $\mathbf{C}_{AC} \neq 0$ \cite{32}, we may follow a similar approach.\\\\
\textit{Usefulness of imaginary projectors.-} Consider a one parameter family of slice states\\
\begin{equation}
\label{eq6}
\ket{\psi_s(a)}=b\ket{000} + b\ket{100} + a\ket{111}
\end{equation}
satisfying $a,b\geqslant 0$ and $a^2 + 2b^2 = 1$. Next we consider a class of generic qubit projective measurement basis : 
\begin{align*}
\ket{+k} &= \cos\alpha\ket{0} + \mathrm{e}^{\mathrm{i}\beta}\sin\alpha\ket{1}\\
\ket{-k} &= \mathrm{e}^{\mathrm{-i}\beta}\sin\alpha\ket{0} - \cos\alpha\ket{1}
\end{align*}
such that $\alpha\in[0,\pi/4]$, $\beta\in[0,\pi)$. The goal here is to concentrate entanglement between $B$ and $C$ with $A$ employing the measurement basis given above. Both the states are imaginary states and infact their imaginarity is quantified by the trace norm $\mathcal{R}(\ket{+k})=\mathcal{R}(\ket{-k})=\sin{2\alpha}\sin{\beta}\neq 0\hspace{1mm}, \forall \beta\in(0,\pi)$. If $A$ now measures her qubit of the state (\ref{eq6}) using the projectors $\{\ket{+k}_{A}\bra{+k},\ket{-k}_{A}\bra{-k}\}$, the average entanglement concentrated between $B$ and $C$ is given by
\begin{align*}
E_{2,Im}^{B|C} =& 2\min \{a^2 \sin^2 \alpha, b^2 (1+\sin2\alpha\cos\beta)\}+\\
& 2\min \{a^2 \cos^2 \alpha, b^2 (1-\sin2\alpha\cos\beta)\}
\end{align*}
For real operation, putting $\beta=0$ we have
\begin{align*}
E_{2,Re}^{B|C} =& 2\min \{a^2 \sin^2 \alpha, b^2 (1+\sin2\alpha)\}+\\
& 2\min \{a^2 \cos^2 \alpha, b^2 (1-\sin2\alpha)\}
\end{align*}
By elementary but somewhat tedious calculations, it can be readily verified that $E_{2,Im}^{B|C}\geqslant E_{2,Re}^{B|C}\hspace{1mm} \forall \alpha\in[0,\pi/4]$.\\
Implementing the two outcome projective measurement $\mathfrak{M}=\{M_{\pm k}\}$ where $M_{\pm k}=\ket{\pm k}_{A}\bra{\pm k}$, is quite straightforward in linear optical setup. The apparatus required to implement such measurements consists of standard half-wave and quarter-wave plates, along with beam splitters. For $\mathfrak{M}$ to be implemented, we only need 11 unset wave plates. This follows from the singular value decomposition of the measurement operators: $M_{\pm k}=U_{\pm k} D_{\pm k} V_{\pm k}^\dagger$ where $U_{\pm k}$, $V_{\pm k}$ are unitaries. $M_{\pm k}$ also satisfies the relation $M_{+k}^\dagger M_{+k}+M_{-k}^\dagger M_{-k}=\mathbb{I}$ implying $V_{-k}D_{-k}^2 V_{-k}^\dagger=\mathbb{I}-V_{+k}D_{+k}^2 V_{+k}^\dagger$ \emph{i.e.}, $V_{-k}D_{-k}^2 V_{-k}^\dagger=V_{+k}(\mathbb{I}-D_{+k}^2)V_{+k}^\dagger$ and therefore, $V_{+k}=V_{-k}$ and $D_{-k}=\mathbb{I}-D_{+k}$. The unitaries $U_{\pm k}, V_{+k}$ can be implemented by three wave plates whereas the diagonal operators $D_{\pm k}$ can be realized by three beam displacers and five wave plates--two of which are unset. Compared to this, any real projectors $\{\ket{\pm k_\alpha}\bra{\pm k_\alpha}\}$ can be realized with only five unset wave plates (more details can be found in Ref. \cite{30,37}).

\section{Three-Qubit Measurement}
\label{sec5}
\begin{figure}[h!]
		\centering
		\includegraphics[scale=0.43]{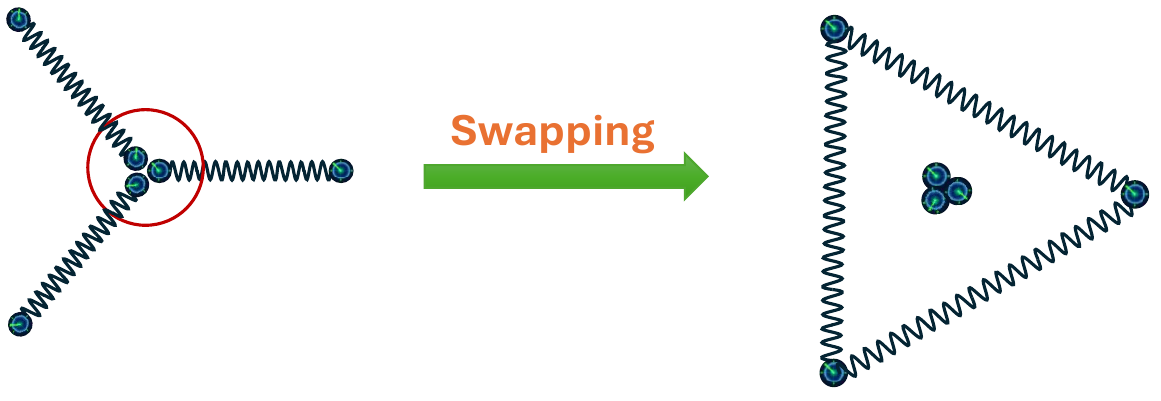}
		\caption{(Color online) \emph{Three-qubit swapping protocol}--the red circle denotes measurement in three-qubit measurement basis. The output state in the right is a three-qubit genuinely entangled state.}
		\label{fig1}
\end{figure}

One of the cornerstone phenomenon in quantum information theory is entanglement swapping \cite{15}. In this scenario $B$ shares a Bell state with $A$, and $C$ also shares a Bell state with $A$. Then $A$ performs a joint measurement in the Bell basis and broadcasts the outcome via a classical channel. As a result, a Bell state is generated between $B$ and $C$, despite the fact that they have never interacted in the past. Even when the shared states are non-maximally entangled pure states, a Bell state can still be generated probabilistically. In a multipartite entanglement-swapping variant, $A$ initially shares Bell pairs with $B$, $C$ and $D$. A three-qubit joint measurement by $A$, followed by classical communication of the outcome, projects $B$, $C$ and $D$ onto a maximally entangled GHZ state.\\
A straightforward structural generalization of Bell state from two-qubit system to three-qubit systems is the GHZ state $\frac{1}{\sqrt{2}}\ket{000+111}$ (upto local unitaries). Thus instead of taking $\{\frac{1}{\sqrt{2}}\ket{00\pm 11}, \frac{1}{\sqrt{2}}\ket{01\pm 10}\}$ as measurement basis, one could take
\begin{equation}
\label{eq7}
\begin{split}
\frac{1}{\sqrt{2}}\ket{000\pm 111}, \frac{1}{\sqrt{2}}\ket{001\pm 110}\\
\frac{1}{\sqrt{2}}\ket{010\pm 101}, \frac{1}{\sqrt{2}}\ket{100\pm 011}
\end{split}
\end{equation}
as an orthonormal measurement basis for the three-qubit case. Measuring in this basis, one will find that corresponding to every outcome, a gGHZ state will be created among the three parties $A$, $B$ and $C$ (see FIG. \ref{fig1}). These gGHZ states can subsequently be converted into the GHZ state (upto local unitaries) with non-zero probabilities \cite{38}. Beyond information processing tasks requiring genuine three-qubit entanglement, one of the most significant applications of genuinely entangled states is their ability to generate Bell states at a later stage. For an arbitrary pure three-qubit state shared among $A$, $B$ and $C$, the maximal average pure state entanglement that can be concentrated between $A$ and $B$ with the assistance from $C$ is given by the quantity $E_2^{A|B}$. If the task of creating a genuinely entangled state is to generate a Bell state later, we wonder if the basis mentioned in (\ref{eq7}) forms the optimal basis. Intriguingly, we find that this is not the case.\\
 Let
\begin{equation}
\label{eq8}
\ket{\phi}=\sqrt{\phi_0}\ket{00}+\sqrt{\phi_1}\ket{11}
\end{equation}
be a non-maximally entangled two qubit state satisfying $\phi_0 > \phi_1 >0$ and $\phi_0+\phi_1 = 1$. Instead of a Bell state, a copy of the state $\ket{\phi}$ be now shared between each of the pairs $A_1\leftrightarrow B$, $A_2\leftrightarrow C$ and $A_3\leftrightarrow D$ where $A_1$, $A_2$ and $A_3$ denotes the three subparties of $A$, each holding the corresponding qubit. Then the joint state of the four parties $A,B,C$ and $D$ is:
\begin{equation*}
\zoom{c}{
\begin{aligned}
\ket{\Phi}_{ABCD}=&\ket{\phi}_{A_1B}\otimes\ket{\phi}_{A_2C}\otimes\ket{\phi}_{A_3D}\\
=&[\phi_0\sqrt{\phi_0}\ket{000,000}+\phi_1\sqrt{\phi_1}\ket{111,111}+\\
&\phi_1\sqrt{\phi_0}(\ket{110,110}+\ket{101,101}+\ket{011,011})+\\
&\phi_0\sqrt{\phi_1}(\ket{100,100}+\ket{010,010}+\ket{001,001})]_{A_1A_2A_3,BCD}
\end{aligned}}
\end{equation*}
First $A$ measures the three-qubits $A_1,A_2$ and $A_3$ in his possession, jointly in the basis (\ref{eq7}) and obtains \cite{39}: 
\begin{equation}
\label{eq9}
E_{2,GHZ}^{A|B}=2\phi_1 ^2 (\phi_1 + 3\phi_0)
\end{equation}
 At this point we seek a new measurement basis that could potentially surpass the GHZ-basis. There are two types of three-qubit genuinely entangled states characterized on their tensor ranks-- 2-rank states belonging to the GHZ class and 3-rank states belonging to the W class. So we argue that only considering one of them as a basis may not be the optimal choice in this situation. A necessary requirement for any basis is that it should be genuinely entangled. Since we want to include the W-class states in the basis and by observing the structure of the state $\ket{\Phi}_{ABCD}$, we consider $\ket{W_1}=\frac{1}{\sqrt{3}}(\ket{100}+\ket{010}+\ket{001})$ as one element of the measurement basis. We also consider the two phase rotated W-class states $\ket{W_2},\ket{W_3}$ (see Eq.(\ref{eq10})) orthonormal to $\ket{W_1}$. Note that after the swapping, our goal is to eventually localize the tripartite entanglement into bipartite entanglement. Thus it is also reasonable to demand for a basis which results in a post-measurement state whose $SCP$ is invariant under permutation of parties. The states $\ket{W_i}$ ($i=1,2,3$) satisfies all these requirements. To complete the basis one could consider $\ket{W_1^\perp}=\frac{1}{\sqrt{3}}(\ket{110}+\ket{101}+\ket{011})$ along with its two phase rotated orthonormal W-class states and the maximally entangled GHZ state $\frac{1}{\sqrt{2}}\ket{000\pm111}$. We find that these eight states form an orthonormal basis and already yield better performance than the measurement strategy in Eq. (\ref{eq7}). We now show that further improvement is possible by allowing complex coefficients in the measurement basis. To this end, we formally introduce the three-qubit GHZ-W (GW) basis, which incorporates imaginary phases and enables enhanced performance: 
\begin{equation}
\label{eq10}
\begin{aligned}
\ket{G_1}=&\frac{1}{\sqrt{5}}(\ket{000}+\ket{110}+\ket{101}+\ket{011}+\ket{111})\\
\ket{G_2}=&\frac{1}{\sqrt{5}}(\ket{000}+\alpha\ket{110}+\alpha^2\ket{101}+\alpha^3\ket{011}+\alpha^4\ket{111})\\
\ket{G_3}=&\frac{1}{\sqrt{5}}(\ket{000}+\alpha^2\ket{110}+\alpha^4\ket{101}+\alpha\ket{011}+\alpha^3\ket{111})\\
\ket{G_4}=&\frac{1}{\sqrt{5}}(\ket{000}+\alpha^3\ket{110}+\alpha\ket{101}+\alpha^4\ket{011}+\alpha^2\ket{111})\\
\ket{G_5}=&\frac{1}{\sqrt{5}}(\ket{000}+\alpha^4\ket{110}+\alpha^3\ket{101}+\alpha^2\ket{011}+\alpha\ket{111})\\
\ket{W_1}=&\frac{1}{\sqrt{3}}(\ket{100}+\ket{010}+\ket{001})\\
\ket{W_2}=&\frac{1}{\sqrt{3}}(\ket{100}+\omega\ket{010}+\omega^2\ket{001})\\
\ket{W_3}=&\frac{1}{\sqrt{3}}(\ket{100}+\omega^2\ket{010}+\omega\ket{001})
\end{aligned}
\end{equation}
where $\omega$ denotes the cube root of unity and $\alpha$ denotes the fifth root of unity. It is quite trivial to observe that $\braket{G_i|W_j}=0 \: \forall i,j$. Further $\braket{G_i|G_j}=\frac{1}{5}\sum_{k=0}^4 \alpha^k=\delta_{ij}$ and $\braket{W_i|W_j}=\frac{1}{3}\sum_{k=0}^2 \omega^k=\delta_{ij}$, where $\delta_{ij}$ denotes the Kronecker delta. It is also starightforward to show that they satisfy the completeness relation, as they are all orthonormal and spans the whole three-qubit ($8$-dimensional) Hilbert space. In terms of imaginarity, except the two states $\ket{G_1}$ and $\ket{W_1}$, all other basis elements are \emph{maximally imaginary} with respect to the measure RoI.\\
Alice now measures her qubits in the GW-basis and gets herself decoupled from the system. The post-measurement genuinely entangled states generated among $B$, $C$ and $D$ are given by:
\begin{equation}
\label{eq11}
\zoom{c}{
\begin{aligned}
&\frac{1}{\sqrt{k}}\{\phi_0\sqrt{\phi_0}\ket{\bf{0}}+\phi_1\sqrt{\phi_0}(\ket{\bf{6}}+\ket{\bf{5}}+\ket{\bf{3}})+\phi_1\sqrt{\phi_1}\ket{\bf{7}}\},\\
&\frac{1}{\sqrt{k}}\{\phi_0\sqrt{\phi_0}\ket{\bf{0}}+\phi_1\sqrt{\phi_0}(\alpha\ket{\bf{6}}+\alpha^2\ket{\bf{5}}+\alpha^3\ket{\bf{3}})+\alpha^4\phi_1\sqrt{\phi_1}\ket{\bf{7}}\},\\
&\frac{1}{\sqrt{k}}\{\phi_0\sqrt{\phi_0}\ket{\bf{0}}+\phi_1\sqrt{\phi_0}(\alpha^2\ket{\bf{6}}+\alpha^4\ket{\bf{5}}+\alpha\ket{\bf{3}})+\alpha^3\phi_1\sqrt{\phi_1}\ket{\bf{7}}\},\\
&\frac{1}{\sqrt{k}}\{\phi_0\sqrt{\phi_0}\ket{\bf{0}}+\phi_1\sqrt{\phi_0}(\alpha^3\ket{\bf{6}}+\alpha\ket{\bf{5}}+\alpha^4\ket{\bf{3}})+\alpha^2\phi_1\sqrt{\phi_1}\ket{\bf{7}}\},\\
&\frac{1}{\sqrt{k}}\{\phi_0\sqrt{\phi_0}\ket{\bf{0}}+\phi_1\sqrt{\phi_0}(\alpha^4\ket{\bf{6}}+\alpha^3\ket{\bf{5}}+\alpha^2\ket{\bf{3}})+\alpha\phi_1\sqrt{\phi_1}\ket{\bf{7}}\}
\end{aligned}
}
\end{equation}
with probabilities $p_1=p_2=p_3=p_4=p_5=k/5$ and each having $SCP=\Bigl(k-\sqrt{k^2-4\phi_0^2\phi_1^2(2-3\phi_0\phi_1)}\Bigr)/k$ and also,
\begin{equation}
\label{eq12}
\begin{aligned}
&\frac{1}{\sqrt{3}}(\ket{\bf{4}}+\ket{\bf{2}}+\ket{\bf{1}}),\\
&\frac{1}{\sqrt{3}}(\ket{\bf{4}}+\omega\ket{\bf{2}}+\omega^2\ket{\bf{1}}),\\
&\frac{1}{\sqrt{3}}(\ket{\bf{4}}+\omega^2\ket{\bf{2}}+\omega\ket{\bf{1}})
\end{aligned}
\end{equation}
with probabilities $p_6=p_7=p_8=\phi_0^2\phi_1$ and $SCP=2/3$, each. Here we have used the shorthand notations $\ket{\bf{0}}:=\ket{000}, \ket{\bf{1}}:=\ket{001}, \ket{\bf{2}}:=\ket{010}, \ket{\bf{3}}:=\ket{011}, \ket{\bf{4}}:=\ket{100}, \ket{\bf{5}}:=\ket{101}, \ket{\bf{6}}:=\ket{110}, \ket{\bf{7}}:=\ket{111}$ for brevity and also, $k=\phi_0^3+\phi_1^3+3\phi_0\phi_1^2$. Observe that using the complex roots of unity has resulted in $SCP$ that is party-permutation invariant
. Thus the average entanglement yield is given by:
\begin{equation}
\zoom{c}{
\label{eq13}
E_{2,GW}^{A|B} = 1-\phi_0 ^2\phi_1-\sqrt{(\phi_0 ^3 + \phi_1 ^3 + 3\phi_0\phi_1 ^2)^2 - 4\phi_0 ^2\phi_1 ^2 (2-3\phi_0\phi_1)}
}
\end{equation}

	
	\begin{figure}[h!]
    \centering
         \subfloat[]{\includegraphics[scale=0.65]{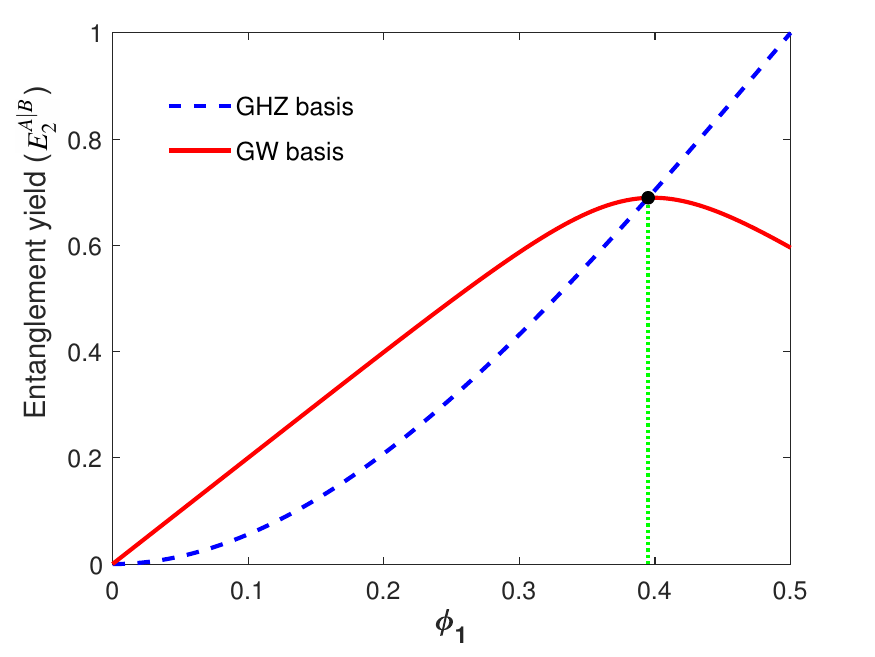} \label{fig2a}}
    \hfill
    \subfloat[]{\includegraphics[scale=0.6]{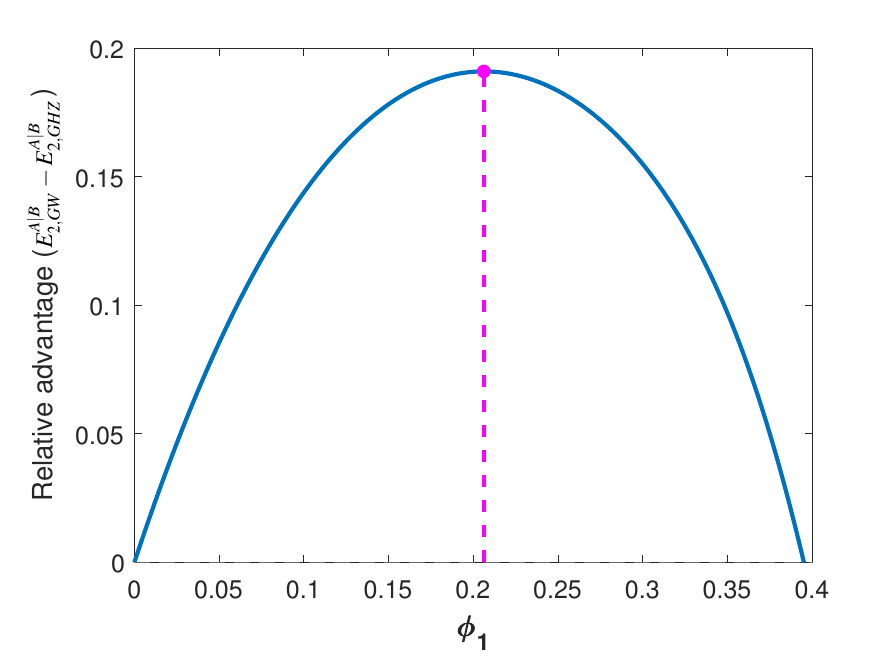} \label{fig2b}}
    \captionsetup{justification=raggedright, singlelinecheck=false}
    \caption{(Color online) (a) The green vertical line incident upon the horizontal $\phi_1$-axis denotes the value of $\phi_1$ where the advantage of GW-basis ceases to exist. $E_{2,GW}^{A|B}\geqslant E_{2,GHZ}^{A|B}$ for $0\leqslant\phi_1\leqslant 0.39493$. (b) The vertical axis denotes the relative advantage of GW basis over GHZ basis \emph{i.e.}, $E_{2,GW}^{A|B}-E_{2,GHZ}^{A|B}$ and the maximum advantage is observed at $\phi_1=0.206$ (incidence point of the magenta line on the $\phi_1$-axis) by $0.191$ ebits.}
    \label{fig2}
\end{figure}

In the FIG. (\ref{fig2a}), we plot the entanglement yield from both GHZ and GW-basis. Clearly for $0\leqslant\phi_1\leqslant 0.39493$, the yield from GW-basis is way higher than the GHZ basis. The actual increment of entanglement is plotted in FIG. (\ref{fig2b}). We have also shown that at $\phi_1=0.206$, the advantage of GW-basis is maximal compared to the GHZ-basis. Thus for three-qubit case, we establish that for certain non-maximally entangled states, the imaginary measurement basis yields higher than measuring in maximally entangled GHZ-basis (\ref{eq7}). In fact if we measure the entanglement of maximally entangled GHZ-basis (\ref{eq7}) in terms the average $SCP$ of the basis elements, then it turns out to be: $\frac{1}{8}\sum_{k=1}^8 SCP(\ket{GHZ_k})=1$. Compared to this, for the GW-basis the $SCP$ is:
\begin{equation}
\label{eq14}
E_2^{A|B}(\ket{G_i})=E_2^{A|C}(\ket{G_i})=E_2^{B|C}(\ket{G_i})=1-\frac{1}{\sqrt{5}}
\end{equation} and
\begin{equation}
\label{eq15}
E_2^{A|B}(\ket{W_j})=E_2^{A|C}(\ket{W_j})=E_2^{B|C}(\ket{W_j})=\frac{2}{3}
\end{equation}
 $\forall i=1,2,3,4,5;j=1,2,3$. Therefore the average is: $\frac{1}{8}\sum_{k=1}^8 SCP(\ket{\psi_k})=\frac{7-\sqrt{5}}{8}\approx 0.6$ where $\ket{\psi_k}\in\{G_i,W_j\}$.\\
Moreover creating the GW-basis experimentally in linear optical setup is not very difficult either. Experimental methods to generate maximally entangled GHZ or W states is quite developed in photonic systems \cite{40,41,42,43,44,45,46}. The set $\{G_i\}$ belongs to the GHZ-SLOCC class and the set $\{W_i\}$ are infact three maximally entangled W states. So after generating one of the three $\ket{W_i}$s, one only needs to apply two local phase rotations in order to obtain the other two $\ket{W_i}$s. As we have also mentioned in Sec. \ref{sec4}, implementing such qubit operations are straightforward in linear optical setup. Any GHZ-SLOCC class state can be expressed as \cite{31,33}:
\begin{equation*}
\ket{\psi_{GHZ}}=\sqrt{K}(\cos{\delta}\ket{a}\ket{b}\ket{c}+\mathrm{e}^{\mathrm{i}\varphi}\sin{\delta}\ket{\phi_A}\ket{\phi_B}\ket{\phi_C})
\end{equation*}
where $\ket{\phi_A}=\cos\theta_1\ket{a} + \sin\theta_1\ket{a^\perp}$, $\ket{\phi_B}=\cos\theta_2\ket{a} + \sin\theta_2\ket{a^\perp}$, $\ket{\phi_C}=\cos\theta_3\ket{a} + \sin\theta_3\ket{a^\perp}$ and $K=(1+2\cos\delta\sin\delta\cos\theta_1\cos\theta_2\cos\theta_3\cos\varphi)^{-1}$. The ranges of the five parameters are: $\delta\in(0,\frac{\pi}{4}]$; $\theta_1, \theta_2, \theta_3\in(0,\frac{\pi}{2}]$; $\varphi\in[0,2\pi)$. Given a three-qubit maximally entangled GHZ state $\ket{GHZ}=\frac{1}{\sqrt{2}}\ket{000+111}$, one could easily obtain any GHZ SLOCC class state $\ket{\psi_{GHZ}}$ through local POVMs $A\otimes B\otimes C$ such that\\
\begin{equation*}
\ket{\psi_{GHZ}}=\frac{1}{\sqrt{p}}(A\otimes B\otimes C)\ket{GHZ}
\end{equation*}
where $A=\cos\delta\ket{a}\bra{a}+\mathrm{e}^{\mathrm{i}\varphi}\sin\delta\cos\theta_1\ket{a}\bra{a^\perp}+\mathrm{e}^{\mathrm{i}\varphi}\sin\delta\sin\theta_1\ket{a^\perp}\bra{a^\perp}$, $B=\ket{b}\bra{b}+\cos\theta_2\ket{b}\bra{b^\perp}+\sin\theta_2\ket{b^\perp}\bra{b^\perp}$ and $C=\ket{c}\bra{c}+\cos\theta_3\ket{c}\bra{c^\perp}+\sin\theta_3\ket{c^\perp}\bra{c^\perp}$ and $p=\bra{GHZ}A^\dagger A\otimes B^\dagger B\otimes C^\dagger C\ket{GHZ}$.\\
For example,
\begin{equation*}
\begin{aligned}
\ket{G_1}=&\frac{1}{\sqrt{5}}(\ket{000}+\ket{110}+\ket{101}+\ket{011}+\ket{111})\\
		=&\cos\delta\ket{a_1b_1c_1}+\sin\delta\ket{a_2b_2c_2}
\end{aligned}
\end{equation*}
where $\cos\delta=\sqrt{\frac{5+\sqrt{5}}{10}}$, $\sin\delta=\sqrt{\frac{5-\sqrt{5}}{10}}$, $\ket{a_1}=\csc\delta[\frac{1}{2}(1-\frac{1}{\sqrt{5}}\ket{0})+\frac{1}{\sqrt{5}}\ket{1}]$, $\ket{a_2}=\sec\delta[\frac{1}{2}(1+\frac{1}{\sqrt{5}}\ket{0})+\frac{1}{\sqrt{5}}\ket{1}]$, $\ket{b_1}=\ket{c_1}=\frac{1}{\sqrt{5}}\sec\delta[\ket{0}+\frac{1+\sqrt{5}}{2}\ket{1}]$ \& $\ket{b_2}=\ket{c_2}=\frac{1}{\sqrt{5}}\csc\delta[\ket{0}+\frac{1-\sqrt{5}}{2}\ket{1}]$. For simplicity first apply the local unitaries $U_A\otimes U_B\otimes U_C$ to convert $\ket{GHZ}$ from computational basis to the basis $\{\ket{a_1},\ket{a_1^\perp}\}_A\otimes\{\ket{b_1},\ket{b_1^\perp}\}_B\otimes\{\ket{c_1},\ket{c_1^\perp}\}_C$, where $U_A=\ketbra{a_1}{0}+\ketbra{a_1^\perp}{1}$, $U_B=\ketbra{b_1}{0}+\ketbra{b_1^\perp}{1}$ \& $U_C=\ketbra{c_1}{0}+\ketbra{c_1^\perp}{1}$. Then one only needs to apply the  local POVM $A\otimes B\otimes C$ on the rotated state to obtain $\ket{G_1}$ where $A=\cos\delta\ketbra{a_1}{a_1}+\sin\delta\cos\theta_1\ketbra{a_1}{a_1^\perp}+\sin\delta\sin\theta_1\ketbra{a_1^\perp}{a_1^\perp}$, $B=C=\mathbb{I}$ and $\cos\theta_1=\frac{2}{\sqrt{5}}$, $\sin\theta_1=\frac{1}{\sqrt{5}}$. The other GHZ SLOCC class states can be generated in a similar way.
\section{Quantum Entanglement Percolation}
\label{sec6}
In this section we show an application of the entanglement swapping protocol with imaginary basis in \enquote*{quantum entanglement percolation} (QEP) protocols \cite{47}.\\
\begin{figure*}[t]
  \centering
  \includegraphics[scale=0.3]{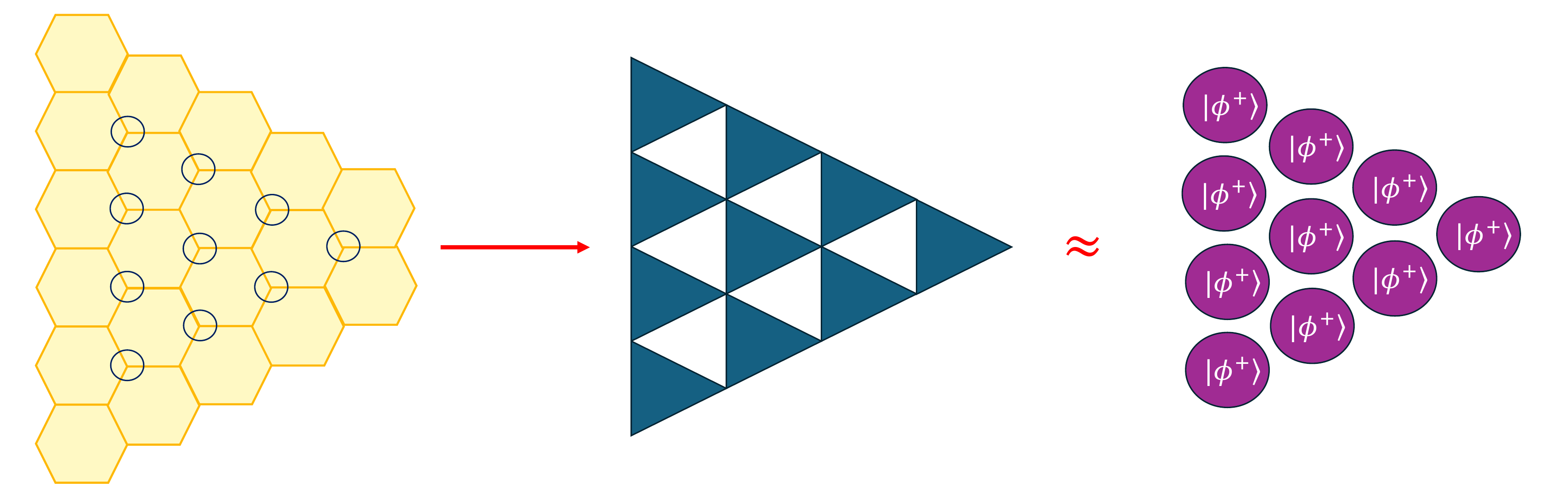}
  \caption{(Color online) The circles on the alternating nodes of the honeycomb lattice in left figure denotes the three-qubit joint measurement by GW-basis. The lattice is then transformed into a triangular lattice where each filled triangle is some genuinely entangled three-qubit state from (\ref{eq11}) or (\ref{eq12}). All of these triangles can be converted into singlets with probability $p_0$ and therefore we consider each triangle as a Bell state. This essentially converts the bond percolation problem into a site percolation problem where each site is represented by a Bell state with occupation probability $p_0$.
  \label{}
  }
  \label{fig3}
\end{figure*}
Bipartite entanglement swapping protocols lies at the heart of distributing entanglement through long distances. For instance, let $A$ holds a two qubit Bell state and wishes to send one of the entangled qubit (photon) to $B$. Then she has to send the qubit through some optical fiber. But the transmittance of an optical fiber decreases exponentially \cite{48}. To overcome this shortcoming, $A$ and $B$ can establish several nodes between them with each node sharing a Bell state with its adjacent node(s). In this manner each intermediary node will possess two qubits and will now be required to perform BSM. This method of repeated entanglement swapping is known as quantum repeater protocol.\\
However Bell states are a costly resource and really difficult to sustain for prolonged period due to environmental noise. Instead, partially entangled states are relatively more noise-resilient to noisy channels \cite{49}, than that of Bell states. In 2007, Ac\'in \emph{et} \emph{al.} \cite{47} formulated classical entanglement percolation (CEP) to show entanglement can be distributed in a quantum network consisting of asymptotically large number of nodes. In this setting, each node is connected to a nearby node via a partially entangled state $\ket{\phi}$. The goal of this protocol is to eventually generate a Bell state between two nodes, irrespective of their distance in the network. Utilising the geometry of the network lattice, they further developed QEP, which is a superior technique than that of CEP.\\
In Ref. \cite{47}, the authors considered a double layered honeycomb lattice and strategically applied two-qubit entanglement swapping. As an outcome, they converted the said lattice into a triangular lattice having lower threshold probability. Nevertheless, depending on the geometry of the lattice, there may exist a possibility of jointly measuring multiple qubits. Perseguers \emph{et} \emph{al.} \cite{50} have considered a QEP strategy by involving multipartite states. By means of Monte-Carlo simulations, they managed to lower threshold probabilities of several well known network lattices. Recently Khanna \emph{et} \emph{al.} \cite{39} considered a scenario of single layer honeycomb lattice, upon which they measured using the basis (\ref{eq7}). This strategy is however no better than the usual bipartite strategy in terms of the yield. But we find that measuring in the basis (\ref{eq10}) proves to be otherwise.\\

\textit{The Protocol.-}  Here we focus on the widely studied hexagonal (honeycomb) lattice, which is particularly appealing since it can be mapped onto a triangular lattice via entanglement swapping. The honeycomb lattice is also very popular due to its applicability in condensed matter physics. We first consider a single layer honeycomb lattice as in \cite{39}. Each node of the lattice consists of the partially entangled state $\ket{\phi}$ and each node of the lattice possess three-qubits, each of which is connected to three neighboring nodes (see FIG. \ref{fig3}). Each of the node thus form a trine structure and therefore if we measure using any three-qubit genuinely entangled basis, the trine open structure converts into a genuinely entangled closed triangle, as illustrated in FIG. \ref{fig1}. Meanwhile all the nodes where measurement is performed become isolated and therefore can be eliminated from the lattice. Thus upon measuring all alternating nodes (with Hamming distance 2) of the honeycomb lattice, the entire lattice is effectively converted into a triangular lattice. The blue circles around the specified nodes of the honeycomb lattice on FIG. \ref{fig3} denote the nodes where the measurement is performed. In the resulting triangular lattice, each elementary triangle corresponds to some three-qubit entangled state belonging to either of (\ref{eq11}) or (\ref{eq12}). Moreover it is apparently cumbersome to calculate the bond percolation threshold for a triangular lattice made of three-qubit genuinely entangled states. To circumvent this difficulty, we instead model the system as a site percolation problem, where each site represents a two-qubit Bell state generated with probability $p_0$, which is given by:
$$
p_0=1-\phi_0 ^2\phi_1-\sqrt{(\phi_0 ^3 + \phi_1 ^3 + 3\phi_0\phi_1 ^2)^2 - 4\phi_0 ^2\phi_1 ^2 (2-3\phi_0\phi_1)}
$$
This can be done since each triangle in the triangular lattice can be converted into a Bell state with probability $p_0$ between any pair of parties. 
Since the site percolation threshold for a triangular lattice is $1/2$ \cite{51}, we have
\begin{equation*}
\begin{aligned}
p_0&>\frac{1}{2}\\
\implies\phi_1&\in(0.252136, 0.5)
\end{aligned}
\end{equation*}
This implies that one could use input partially entangled states with entanglement as low as $S_{GW}\approx 0.8146$ ebits for successful percolation, where $S_{GW}$ denotes the Von-Neumann entropy of the reduced state of $\ket{\phi}$. This requirement is substantially smaller than that of the GHZ basis, which demands a minimum of $S_{GHZ}\approx 0.9112$ ebits of entanglement. In terms of the bond occupation probability, the earlier strategy \cite{39} with GHZ basis requires a minimun probability $0.6527$ in each bond of the honeycomb lattice. But measuring in GW-basis reduces the requirement of this probability to $0.5043$ ($22.7\%$ reduction).
 Further to connect any two node in the triangular lattice, one needs to design a path between two nodes and perform appropriate local measurements to generate Bell states along the path so that usual perfect entanglement-swapping protocol can be implemented.\\
 Intriguingly, two distinct resources are at play in this scenario--entanglement and imaginarity. The conventional strategy based on the GHZ basis employs maximal entanglement but relies on a purely \emph{real} measurement basis. In contrast, the GW basis requires a smaller amount of entanglement, albeit at the cost of a higher degree of imaginarity (see TABLE \ref{table1}). This reveals a clear trade-off between these two resources, which ultimately leads to an enhanced success probability for the task under consideration.\\
%
%

\begin{table}[h]
\centering
\begin{tabular}{ |p{2cm}||p{1.6cm}|p{1.6cm}|p{2cm}| }
 \hline
 \multicolumn{4}{|c|}{Entanglement vs Imaginarity} \\
 \hline
 Measurement basis & Entanglem- ent & Imaginarity & Bond occupation probability \\
 \hline
 GHZ basis \cite{39} & 1 & 0 & $p_{GHZ}=p_{CEP}$ \\
 \hline
 GW basis & 0.6 $\downarrow$ & 0.75 $\uparrow$ & $p_{GW} \downarrow$ \\
 \hline
\end{tabular}

\caption{The table compares the two types resource employed in the three-qubit measurement bases-- entanglement and imaginarity. The entanglement content is quantified by computing the average $SCP$ over all basis elements. Likewise, the imaginarity is quantified by taking the average RoI of all the basis elements with respect to the computational basis. Since measuring in GHZ basis cannot do any better than than CEP, the bond occupation probability of both CEP ($p_{CEP}$) and QEP with GHZ basis are identical ($p_{GHZ}$). However measuring in GW basis reduces the occupation probability ($p_{GW}$) of input states.}
\label{table1}
\end{table}

\section{Discussion}
\label{sec7}
Whether quantum theory can be described in terms of only real numbers is a long standing problem. Renou \emph{et. al} \cite{52} provided a strong argument in favor of the use of complex numbers. However recent arguments \cite{53,54,55,56} indicates that the debate is perhaps not settled yet. Nevertheless, everybody seems to agree on the fact that the use of complex numbers in quantum theory is anyway a convenient tool. In fact based on the fact that quantum states with complex coefficients requires more physical resources than real ones, a new resource theory of imaginarity has been formulated in recent times.\\
Little is explored about the resources that are used in measurement setups. However, there is an increasing interest to characterize the resources, viz., entanglement used in measurement. We begin our work by showing that genuine entanglement of a pure three-qubit real slice state--an otherwise fascinating class of state, can be concentrated in a bipartition much easily if one uses maximally imaginary states as measurement basis. In particular, instead of real bases, if one uses $\ket{\hat{\pm}}$ basis, then it is only required to know $i,j$ such that $C_{ij}\neq0$. This greatly reduces the need for local quantum state tomography in order to precisely measure the parameters of the state. Further, we consider a one parameter class of the said real slice states and show that imaginary projectors outperform real ones.\\
Next, we consider entanglement swapping protocols that involve more than one party. Here we have specifically considered an entanglemet swapping protocol involving three-qubit joint measurement. We defined GW-basis which substantially outperforms the maximally entangled GHZ-measurement basis. This is yet another example where a non-standard, imaginary non-maximally entangled measurement basis is shown to yield more entanglement than the standard maximally entangled basis. These results also indicates the intrinsic structural richness of the three-qubit systems compared to two-qubit. Interestingly, the advantage of GW-basis over GHZ-basis manifests prominently for relatively weakly entangled input states of the form: $\ket{\phi}=\sqrt{1-\phi_1}\ket{00}+\sqrt{\phi_1}\ket{11}$ with $0\leqslant\phi_1\leqslant 0.39493$. This further reveals operational advantage of entanglement dilution in quantum information processing tasks, as has already been argued in \cite{49}. It is also worth noting that at the first glance, the GW-basis is fairly asymmetric, unlike iso-entangled bases of two qubits \cite{22}. But the GHZ and W class elements in the basis are separately iso-entangled in the sense that each state possesses identical $SCP$s irrespective of the bipartitions, \emph{i.e.},
\begin{equation*}
E_2^{A|B}(\ket{\psi})=E_2^{A|C}(\ket{\psi})=E_2^{B|C}(\ket{\psi})
\end{equation*}
 $\forall \ket{\psi}\in\{\ket{G_i},\ket{W_j}:i=1,2,3,4,5;j=1,2,3\}$ and 
 \begin{equation*}
 \begin{aligned}
&E_2^{A|B}(\ket{G_k})=E_2^{A|B}(\ket{G_l}) \;\; \forall k,l\\
&E_2^{A|B}(\ket{W_m})=E_2^{A|B}(\ket{W_n}) \;\; \forall m,n
\end{aligned}
\end{equation*}
This multipartite swapping protocol admits a direct application to quantum entanglement percolation. We find that need for entanglement is reduced by approximately ($\Delta_S=S_{GHZ}-S_{GW}\approx$) $0.1$ ebit or ($\frac{\Delta_S}{S_{GHZ}}\times100\%\approx$) $10.6\%$. We also find that bond occupation probability for honeycomb lattice is reduced by a staggering $22.7\%$. Compared to the usual maximally entangled GHZ measurement basis, we also observe a trade off of entanglement and imaginarity in the GW basis. We see that the need for entanglement is lowered but the requirement of imaginarity has increased from 0 to 0.75.\\
Furthermore these three-qubit measurement basis states are also not very diificult to generate experimentally in usual linear optical systems. The three-qubit measurement basis states can be generated from maximally entangled GHZ or W staes with local measurements. Nevertheless, identifying an optimal three-qubit measurement basis requires optimization over a high-dimensional parameter space, and whether the GW basis is truly optimal remains an open question. The GW-basis does not depend on the parameters values of $\ket{\phi}$. One could try to construct measurement basis keeping in mind some fixed parameter values of $\ket{\phi}$ --which may perform even better than GW-basis.\\
Pimpel, Renner and Tavakoli \cite{57} has recently investigated the problem of generating a full iso-entangled basis by applying local uniatries in bipartite as well as multipartite qudit systems. For three qubit systems, the existence of any such basis necessarily implies that the basis either belong to GHZ or W class. Therefore the basis will never be able to utilise the full entanglement structure available in the pure three qubit state space. On the contrary, the GW basis contains both GHZ as well as W class states, and the basis, although not LU equivalent, but has the property of being iso-entangled in terms of SCP, separately in both SLOCC classes [see Eq. (\ref{eq14}) \& (\ref{eq15})]. Another open problem raised in the same work is regarding the resilience of the measurement basis under particle loss. Construction of such basis is a step toward building noise-resilient entanglement swapping protocols. Unlike the GHZ basis, the GW basis is quite resilient since $\mathbf{C}_{ij}\neq 0\: \forall i,j$, of the states $\{\ket{G_k},\ket{W_m}\}_{k>1}$. Thus, our construction of GW basis also partially addresses the open problems raised in \cite{57}.\\
Moreover it has been shown that two \emph{real} orthonormal Bell-diagonal mixed states cannot be locally discriminated if one has access to only real operations \cite{24}. Recently it has been shown that for entanglement swapping with certain two-qubit noisy states, using non-maximally entangled measurements could be more useful than BSM in terms of teleportation fidelity \cite{58}. It will be interesting to find whether imaginarity has a role to play in such scenarios for two-qubit systems.  Unfortunately for the mixed states of three-qubit systems, the optimal value of the entanglement of assistance is not known. It would be interesting to find \emph{real} mixed three-qubit states where the imaginarity again proves to be a resource.

\section*{ACKNOWLEDGEMENT}
I.B. acknowledges support from UGC, India. S.B. acknowledges support from CSIR, India. D.S. and I.C. acknowledge that this work is part of the DST-FIST program, India. Both I.B. and S.B. also thank the Harish-Chandra Research Institute, Prayagraj, for their kind hospitality, where part of this work was carried out.

\end{document}